\documentclass [12pt]{article}
\def\maj#1{\ifmmode\mbox{\usefont{U}{msb}{m}{n}#1}\else{\usefont{U}{msb}{m}{n}#1}\fi}
\def\v#1{\mathbf{#1}}
\usepackage{graphics,epsfig,graphicx}
\usepackage{color}

\begin{document}

\title{\textbf{Faraday rotation in photoexcited semiconductors:\\ an 
excitonic many-body effect}}
\author{M. Combescot and O. Betbeder-Matibet
 \\ \small{\textit{Institut des NanoSciences de Paris,}}\\
\small{\textit{Universit\'e Pierre et Marie Curie and Universit\'e Denis
Diderot, CNRS,}}\\ \small{\textit{Campus Boucicaut, 140 rue de
Lourmel, 75015 Paris, France}}}
\date{}
\maketitle

\vspace{0.5cm}

PACS.71.35.-y -- Excitons and related phenomena.

\vspace{2cm}

\begin{abstract}
This letter assigns the Faraday rotation in photoexcited semiconductors to
``Pauli interactions'', \emph{i}.\ \emph{e}., carrier exchanges,
between the real excitons present in the sample and the virtual excitons coupled to
the $\sigma_{\pm}$ parts of a linearly polarized light. While
\emph{direct Coulomb} interactions scatter bright excitons 
into bright excitons, whatever their spins are, \emph{Pauli}
interactions do it for bright excitons \emph{with same spin
only}. This makes these Pauli interactions entirely responsible
for the refractive index difference, which comes from processes in which the 
virtual exciton which is created and the one which recombines are formed with
different carriers. To write this difference in terms of photon
detuning and exciton density, we use our new many-body theory for interacting
excitons. Its multiarm ``Shiva'' diagrams for $N$-body exchanges make transparent the
physics involved in the various terms. This   work also shows the
interesting link which exists between Faraday rotation and  the exciton optical Stark
effect.
\end{abstract}

It is known since a long time that the polarization plane of a linear
light rotates when passing through an optically active medium.
This effect, known as Faraday rotation, is usually explained
rather phenomenologically, by saying that, due to a dissymmetry in the sample ---
which can preexist or be induced --- the
$\sigma_{\pm}$ components of the light have different refractive indices $n_{\pm}$,
the rotation of the polarization plane being proportional to
$(n_+-n_-)$. From a microscopic point of view,
this difference has to come from difference in the interactions of the
sample with the virtual excitations coupled to the unabsorbed $\sigma_{\pm}$
photons. Faraday rotation is thus a powerful tool to
study not only these interactions, but also
the excited state relaxations, in particular, spin relaxation
crucial for spintronics [1-6]. In order to fully
control these studies, a precise microscopic understanding of this rotation is
however highly desirable. 

In the case of semiconductors, the matter excitations
are the excitons. With respect to their possible interactions, doped and
photoexcited samples are very different. Indeed, due to Pauli exclusion, the free
carriers of a doped sample have different energies; this makes their possible
interactions with the virtual excitons coupled to photons rather tricky. Using
experimental results on Faraday rotation and circular dichroism, we have recently
suggested [7] that these carriers do not form trions, as commonly said, but
an intrinsically wide many-body object, singular at threshold. Photoexcited samples
are, in this respect, simpler because the excitons they contain have an
energy which is essentially constant for very heavy holes. However,
even in this case, a fully microscopic theory of Faraday rotation has not been easy
to produce because interactions \emph{between} excitons are difficult to handle
properly due to the exciton composite nature --- which is here crucial.

Our new many-body theory for composite bosons [8] now provides a quite efficient tool
to face such a problem. The diagrammatic representation we have recently proposed
[9], with its multiarm ``Shiva'' diagrams [10] to visualize $N$-body exchanges, allows
a keen understanding of the physics involved in the various terms.

\section{Microscopic approach}

A microscopic description of Faraday rotation goes through the determination of the
linear response function $S_\sigma$ of the material to a probe beam with circular
polarization $\sigma=\pm 1$. This response function, for a matter state
$|\psi\rangle$, reads [11]
\begin{equation}
S_\sigma=\langle\psi|U_\sigma\,\frac{1}{\omega+\mathcal{E}-H+
iO_+}\,U_\sigma^\dag|\psi\rangle\ ,
\end{equation}
where $\omega$ is the photon energy, $\mathcal{E}$ the state $|\psi\rangle$ energy
and $H$ the matter Hamiltonian. For semiconductors, the matter coupling to $\sigma$
photons can be written as $U_\sigma^\dag=\sum_I\delta_{S_i,\sigma}\mu_i^\ast
B_I^\dag$, where $B_I^\dag$ creates an exciton $I=(i,S_i)$ with spin $S_i$, in a
state $i=(\nu_i,\v Q_i)$, while $|\mu_i|$ is the $i$ exciton Rabi energy [12]. The
refractive index difference is related to this response function through
$n_+-n_-\simeq \mathrm{Re}(S_{+1}-S_{-1})/\mathcal{E}_p$, where $\mathcal{E}_p$ is an
energy like constant [13].

\section{Physical origin}

Let us concentrate on quantum wells previously excited by $\sigma_+$ pump photons
tuned on the ground state exciton. Before spin relaxation, this sample
essentially contains
$N$ ground state excitons $(S=1)$, made of (3/2) hole and (-1/2) electron, the
corresponding matter state being [14]
\begin{equation}
|\psi\rangle\simeq [N!F_N]^{-1/2}\,B_O^{\dag N}|v\rangle\ ,
\end{equation}
where $B_O^\dag$ creates an exciton $O=(\nu_o,\v Q_o, S_o=1)$ in the $\nu_o$
relative motion ground state, the exciton momentum $\v Q_o$ being essentially the
pump photon momentum. $N!F_N=\langle v|B_O^NB_O^{\dag N}|v\rangle$ is a normalization
factor which differs from $N!$ due to the exciton composite nature [15].

In $S_\sigma$, the operator
$U_\sigma^\dag$ adds a virtual exciton $I$ to these $N$ excitons $O$, while
$U_\sigma$ destroys a virtual exciton $I'$, \emph{a priori} different from $I$:
As photons interact with a semiconductor through the virtual excitons to
which they are coupled, the response function comes from  the interactions of 
$I$ with the
excitons present in the sample. The exciton
$I'$  which recombines thus results from all the interactions between $I$ and
$|\psi\rangle$ which leave $|\psi\rangle$ unchanged.

If we now consider spins, the exciton $I'$ restoring a
$\sigma_{\pm}$ photon must have a $(\pm 3/2)$ hole and a $(\mp 1/2)$
electron. As fermions are indistinguishable, these carriers can be either the
ones of $I$ or any other similar carriers.
The physical reason for $S_{+1}\neq S_{-1}$ then becomes transparent: If the
sample contains $N$ holes $(3/2)$ and $N$ electrons $(-1/2)$, the virtual exciton
$I'$ in $S_{-1}$, which regenerates a $\sigma_-$ photon, can only be made of the
$(-3/2,1/2)$ carriers of $I$. On the opposite,
$I'$ in $S_{+1}$, which regenerates a $\sigma_+$ photon, can be
made either of the $I$ carriers or of any of the $N$ holes $(3/2)$ and the
$N$ electrons
$(-1/2)$ present in the sample. The processes in which $I'$ is made of carriers
different from $I$ --- one of the two carriers being possibly the same --- are thus
the ones producing $S_{+1}$ different from $S_{-1}$.

To go further and identify these processes precisely,
we make use of the two elementary scatterings of our many-body theory for composite
excitons, namely the (energy-like)
direct Coulomb scatterings, in which the
``in'' and ``out'' excitons keep their carriers, and the (dimensionless)
Pauli scatterings in which the excitons exchange their carriers without
Coulomb.

Due to spin conservation, Pauli scatterings scatter bright
excitons into bright excitons if they have the same spin only, while (direct) Coulomb
scatterings do it whatever their relative spins are. Consequently, as the exciton
$I'$ has to be bright to restore a photon, carrier exchanges through Pauli
scatterings between $I$ and $O$ can produce a bright $I'$ if $I$ and $O$ have same
spin only, \emph{i}.\ \emph{e}., if the pump and the probe have the same circular
polarization.

\section{Dependences in detuning and exciton density}

Let us now feel the physics which controls the
dependence of $(S_{+1}-S_{-1})$ on the two parameters of the problem, namely the
probe photon detuning $\Omega=E_g-\omega$, and the
density
$N/L^D$ of excitons in a sample of size $L$ in $D$ dimension. These parameters can
be associated to the  dimensionless quantities
\begin{equation}
\gamma=R_X/\Omega\ ,\hspace{2cm} \eta=N(a_X/L)^D\ ,
\end{equation}
where $(R_X,a_X)$ are the 3D exciton Rydberg and Bohr radius.

The response function $S_{\sigma}$ defined in eq.(1) can be represented by the
diagram of fig.1a. In the ``box'', any Pauli or Coulomb interaction between
$I$ and the $N$ excitons $O$ can take place, provided that the electron and
hole spins of
$I'$ are the ones of $I$, to restore the $\sigma$ photon.

The $\gamma$ dependence of $S_\sigma$ can be understood
through dimensional arguments. In view of eq.(1), $S_\sigma$ \emph{a priori}
contains two couplings, $U_\sigma$ and $U_\sigma^\dag$, \emph{i}.\
\emph{e}., two Rabi energies, and one energy denominator, of the order of the
detuning. If the excitons in the ``box'' interact via Coulomb scatterings, other
energy denominators, \emph{i}.\
\emph{e}., detunings, must appear to compensate the Coulomb scattering
dimension, while this is unnecessary if they interact through the
dimensionless Pauli scatterings. Consequently,
the $\gamma$ leading term comes from processes with zero Coulomb interaction.

We now turn to the density dependence. Processes in which the exciton $I$ interacts
with \emph{one} exciton $O$, must appear with a factor $N$, as there is $N$
ways to choose this exciton among the $N$ excitons $O$.
These processes produce the $\eta$ linear terms of $S_\sigma$. In the same way, 
processes in which $I$ interacts with \emph{two} excitons $O$ produce the 
$\eta^2$ terms, as there are 
$N(N-1)$ ways to choose two excitons among $N$. And so on \ldots

Let us now visualize these various processes, using our multiarm ``Shiva'' diagrams
for $N$-body exchanges:

(i) The $\gamma$ leading term, linear in $\eta$, is due to Pauli scatterings
between $I$ and $p=1$ exciton $O$ (see fig.1b). In these processes, the
excitons $I'$ and $I$ have one common carrier. Obviously, the pump and the probe must
have the same  circular polarization for these diagrams to exist.

(ii) The $\gamma$ leading term, quadratic in $\eta$, is due to Pauli scatterings
between $I$ and $p=2$ excitons $O$ (see fig.1c). In the two first diagrams, 
$I'$ and $I$ have one common carrier, while in the last diagram,
their two carriers are different. And so on\ldots for the $\eta^p$ term. 

(iii) The next order term in $\gamma$ comes from processes with \emph{one}
Coulomb interaction between $I$ and the excitons $O$. In the $\eta$ term, 
only one of these $O$ excitons enters. It is constructed on the diagrams of fig.1b,
with the Coulomb scattering between any two exciton lines. The $\eta^2$ term is
constructed in a similar way on the diagrams of
fig.1c. And so on \ldots

We could think that processes with one Coulomb scattering, \emph{i}.\ \emph{e}., two
detuning denominators, should behave as
$\gamma^2$. This is more subtle. The virtual excitons coupled to probe
photons are not necessarily ground state excitons, for no energy conservation is
required for virtual processes. It turns out that the
exciton extended states give a singular contribution to $S_\sigma$ which, at large
detuning, transforms the na\"{\i}ve
$\gamma^2$ dependence into
$\gamma^{3/2}$. A way to understand it, is to say that counting processes
with 0, 1, 2\ldots Coulomb interactions amounts to perform an
expansion in $e^2$, proportional to $\gamma^{1/2}$.

By calculating precisely all these contributions, we eventually find that the
expansion of the refractive index difference in terms of $\gamma$ and $\eta$ reads,
for a 2D structure,
\begin{equation}
n_+-n_-\simeq f\,\left[
\gamma\{2\eta-(4\pi/5)\eta^2+O(\eta^3)\}+\gamma^{3/2}\{3\pi\eta+O(\eta^2)\}+O(\gamma^2)
\right]\ ,
\end{equation}
where $f=|\mu|^2/\mathcal{E}'_pR_X$ is a dimensionless constant, with
$\mathcal{E}'_p=\mathcal{E}_p(a_X/L)^D$ and $|\mu|$ being energies free of sample
size (see [12] and [13]). 

\section{Main steps of the theory}

This understanding may appear as wishful thinkings. It
actually follows from our many-body theory for composite excitons.

\noindent(i) The Coulomb expansion of $S_\sigma$, is obtained by passing the 
$B_I^\dag$'s of $U_\sigma^\dag$ over the Hamiltonian, through [8]
\begin{equation}
(a-H)^{-1}B_I^\dag=[B_I^\dag+(a-H)^{-1}V_I^\dag](a-H-E_i)^{-1}\ ,
\end{equation}
which follows from 
$V_I^\dag=[H,B_I^\dag]-E_iB_I^\dag$, where $E_i$ is the $I$ exciton energy.

This leads to split $S_\sigma$ as
$S_\sigma^{(0)}+S_\sigma^{(1)}+S_\sigma^{\mathrm{(corr)}}$, where $S_\sigma^{(0,1)}$
are zero and first order in the ``creation Coulomb potential'' $V_I^\dag$, while
$S_\sigma^{\mathrm{(corr)}}$ contains all higher order terms:
\begin{equation}
S_\sigma^{(0)}=\sum_{I',I}\frac{\delta_{S_{i'},\sigma}\,
\delta_{S_i,\sigma}}{2\Delta_i}\,\left[
\mu_{i'}\mu_i^\ast\,\langle\psi|B_{I'}\,B_I^\dag|\psi\rangle\,+c.c.
\right]\ ,
\end{equation}
\begin{equation}
S_\sigma^{(1)}=\sum_{I',I}\frac{\delta_{S_{i'},\sigma}\,
\delta_{S_i,\sigma}}{2\Delta_{i'}\Delta_i}\,\left[\mu_{i'}\mu_i^\ast\,
\langle\psi|B_{I'}V_I^\dag|\psi\rangle\,+c.c.\right]\ ,
\end{equation}
where $\Delta_i=(-\Omega_i+i0_+)$, with $\Omega_i=E_i-\omega$ being the $I$ exciton
detuning. 

\noindent (ii) In $S_\sigma^{(0)}$, appear the scalar products of $(N+1)$ excitons.
They are calculated using
\begin{equation}
\left[B_M,B_O^{\dag N}\right]=N\,B_O^{\dag N-1}\,(\delta_{M,O}
-D_{MO})-N(N-1)\sum_P\Lambda_h\left(_M^P\ _O^O\right)\,B_P^\dag
B_O^{\dag N-2}\ ,
\end{equation}
which is a generalization for $N>1$, of the equation which defines the
``deviation-from-boson'' operator $D_{MO}$ [8]. This operator is then eliminated
through
\begin{equation}
\left[D_{MI},B_J^\dag\right]=\sum_P \left[\Lambda_h\left(_M^P\
_I^J\right)+\Lambda_h\left(_P^M\ _I^J\right)\right]\,B_P^\dag\ ,
\end{equation}
where the Pauli scattering $\Lambda_h\left(_M^P\ _I^J\right)$ corresponds to a hole
exchange between $I$ and $J$.

For excitons $(I',I)$ coupled to the probe and excitons $O$ created by a pump with
different wavelength, \emph{i}.\ \emph{e}., $\v Q_{i'}=\v Q_i\neq\v Q_o$, the
scalar products of interest [16] in $S_\sigma^{(0)}$ reduce to
\begin{eqnarray}
\langle v|B_O^NB_{I'}B_I^\dag B_O^{\dag N}
|v\rangle=\delta_{S_{i'},S_i}\left\{A_N\delta_{i',i} +\delta_{S_i,S_o}
\left[-A_{N-1}N^2\,I_2^{(0)}(i',i)\right.\right.\nonumber\\
\left.\left.+A_{N-2}N^2(N-1)^2\,I_3^{(0)}(i',i)+\cdots\right]\right\}\
,
\end{eqnarray}
with $A_N=N!F_N$. The processes corresponding to $I_{2,3}^{(0)}$ are the
carrier exchanges of figs.1b,c. For a $\sigma_+$ pump, this leads to
$S_{-1}^{(0)}=\sum_i|\mu_i|^2/\Delta_i$, while
\begin{equation}
S_{+1}^{(0)}-S_{-1}^{(0)}=\sum_{i'i}\frac{1}{2\Delta_i}\,\left\{
\mu_{i'}\mu_i^\ast\left[-N\frac{F_{N-1}}{F_N}\,I_2^{(0)}(i',i)
+N(N-1)\frac{F_{N-2}}{F_N}\,I_3^{(0)}(i',i)+\cdots\right]\
+c.c.\right\}\ .
\end{equation}

\noindent (iii) To calculate the first order term in Coulomb
scatterings $S_\sigma^{(1)}$, we use
\begin{equation}
\langle v|B_O^NB_{I'}V_I^\dag B_O^{\dag N}|v\rangle=N\sum_{JK}
\Xi\left(_J^K\ _I^O\right)\langle v|B_O^NB_{I'}B_J^\dag
B_K^\dag B_O^{\dag N-1}|v\rangle\ ,
\end{equation}
which results from the generalization for $N>1$ excitons of the equation which
defines the (direct) Coulomb scatterings, namely
\begin{equation}
\left[V_I^\dag,B_O^{\dag N}\right]=N\,\sum_{JK}\Xi
\left(_J^K\ _I^O\right)\,B_J^\dag B_K^\dag\,B_O^{\dag N-1}\ .
\end{equation}

The remaining scalar products of $(N+1)$ excitons are then calculated in the same way
as for $S_\sigma^{(0)}$. We end with 
\begin{equation}
S_{+1}^{(1)}-S_{-1}^{(1)}=\sum_{i'i}\frac{1}{2\Delta_{i'}\Delta_i}\,\left\{
\mu_{i'}\mu_i^\ast\left[N\frac{F_{N-1}}{F_N}\,I_2^{(1)}(i',i)
-N(N-1)\frac{F_{N-2}}{F_N}\,I_3^{(1)}(i',i)+\cdots\right] \
+c.c.\right\}\ .
\end{equation}

The processes of $I_2^{(1)}(i',i)$ are the ones of $I_2^{(0)}(i',i)$ with the
Coulomb scattering \emph{before} the carrier exchange, while in the ones of
its complex conjugate, it is \emph{after} the exchange.

\noindent (iv) By using $F_{N-p}/F_N\simeq
(F_{N-1}/F_N)^p$ for $p\ll N$, with $F_{N-1}/F_N\simeq 1+
4\pi\eta/5+O(\eta^2)$ in 2D [15], eqs.(11,14) lead to
\begin{equation}
S_{+1}-S_{-1}\simeq \left(L/a_X\right)^D|\mu|^2\left\{\eta\,
a(\omega)+\eta^2\left[\frac{4\pi}{5}a(\omega)-b
(\omega)\right]+O(\eta^3)\right\}\ ,
\end{equation} 
where $a(\omega)$ precisely reads
\begin{equation}
a(\omega)=\frac{1}{2}\sum_{i'i}\frac{\mu_{i'}\mu_i^\ast}{|\mu|^2}
\left(\frac{I_2^{(0)}(i',i)}{\Omega_i}+\frac{I_2^{(1)}(i',i)}
{\Omega_{i'}\Omega_i}+\cdots\right)\ +c.c.\  =
a^{(0)}(\omega)+a^{(1)}\omega)+\cdots\ ,
\end{equation}
with $\left(I_2^{(0)}(i',i),\,I_2^{(1)}(i',i)\right)$ replaced by
$(L/a_X)^D\left(I_3^{(0)}(i',i),\,I_3^{(1)}(i',i)\right)$ to get $b(\omega)$. 
By performing the sums over $i$ in $a(\omega)$ through closure relations,
we find 
\begin{equation}
a^{(0)}(\omega)=\sum_{\v k}G_{\v k}(\omega)|\langle\v k|\nu_o
\rangle|^2\ +c.c.\ ,
\end{equation}
\begin{equation}
a^{(1)}(\omega)=\sum_{\v k,\v k'}V_{\v k'-\v k}\langle \v k|\nu_o
\rangle\,\left[\langle\nu_o|\v k\rangle-\langle\nu_o|\v k'
\rangle\right]G_{\v k'}(\omega)\left[G_{\v k}^\ast(\omega) -G_{\v
k'}^\ast (\omega)\right]\ +c.c.\ ,
\end{equation}
where $G_{\v k}(\omega)=\sum_{\v k'}\langle\v k|(h+E_g-\omega)^{-1}|
\v k'\rangle$, with $h$ being the exciton Hamiltonian.

These two sums already appeared in our old work on the exciton optical
Stark effect (eqs.\ (6.15,30) of ref.\ [17]). In 2D, they read
$a^{(0)}(\omega)=\Omega^{-1}[2+2\pi\gamma^{1/2}+O(\gamma)]$ and
$a^{(1)}(\omega)=\Omega^{-1}[\pi\gamma^{1/2}+O(\gamma)]$ --- 
while the correlation term would behave as
$\Omega^{-1}O(\gamma)$ [18].

It is, after all, not so surprising to find that the $\eta$
terms of $(S_{+1}-S_{-1})$, \emph{i}.\ \emph{e}., $(n_+-n_-)$, 
are the ones we found in the exciton optical Stark shift,
because  the physics is the same: The exciton optical Stark shift, at
\emph{lowest order in pump intensity}, results from the interactions
of \emph{one} virtual exciton coupled to the unabsorbed pump 
and \emph{one} real exciton created by
the absorbed probe. In the same way, the $\eta$ term of $(S_{+1}-S_{-1})$ results
from the interactions of two excitons: $I$
coupled to the probe and \emph{one} of the $N$ excitons $O$.

\emph{The important step our many-body theory now offers} is the
possibility to go beyond 
linear effects, by considering the interactions of more than two excitons.
In $b(\omega)$, $I$ and two excitons $O$ are involved. Its large
detuning contribution, $b^{(0)}(\omega)$, constructed on
$I_3^{(0)}(i',i)$ in a simlar way as for $a^{(0)}(\omega)$, 
reads
\begin{equation}
b^{(0)}(\omega)=\frac{3}{2}(L/a_X)^2\sum_{\v k}G_{\v
k}(\omega)|\langle \v k|\nu_o\rangle|^4\ +c.c.\ \simeq \Omega^{-1}
[12\pi/5+O(\gamma^{1/2})]\ ,
\end{equation}
while $b^{(1)}(\omega)$ behaves as $\Omega^{-1}O(\gamma^{1/2})$.

By inserting these 2D values of $a(\omega)$ and $b(\omega)$ into eq.(15), we get the
expansion of $(n_+-n_-)$ given in eq.(4).

\section{State of the art}

We have found only one microscopic theory of Faraday rotation in photoexcited
quantum wells. Using 
a previous work on $\chi^{(3)}$ [19], L. Sham and coworkers
[1] have calculated the time evolution of Faraday rotation as
a function of the pump-probe delay,  
with a sample irradiated by circular or linear pumps. 
Using a phenomenological dephasing rate, they find that Faraday rotation decays
smoothly for a circular pump, while a beating arises when the pump
is linear. If, in $B_0^{\dag N}$, we replace
$B_0^\dag$ by a combination of $(\pm 1)$ excitons, the nonlinear
many-body problem we here face becomes far more complicated. This is why we 
stayed with a circular pump. Ref.[1] avoids to face this many-body problem by keeping
terms with
\emph{one} pump exciton, not $N$ as we do. The contribution of
high energy exciton bound and unbound states is also neglected. This is highly
questionable even at small detuning: While, for $\Omega_0$ small, it is easy to
replace
$G_{\v k}(\omega)$ by
$L\Omega_0^{-1}\langle \v k|\nu_0\rangle\langle\nu_0|\v r=\v
0\rangle$ in eqs.\ (17-19), the correlation term $S_\sigma^\mathrm{(corr)}$ is no
more negligible in this limit. There is however \emph{no way} to calculate 
$S_\sigma^\mathrm{(corr)}$ reliably,
except at large detuning [18], or when it is controlled by the biexciton,
\emph{i}.\
\emph{e}., when the exciton optical Stark shift turns from blue to red [20].

\section{Conclusion}

The main goal of this letter is to provide a physical understanding of the
processes producing Faraday rotation in photoexcited semiconductors. This
understanding heavily relies on the concepts of our new many-body theory for composite
bosons, appropriate to face problems in which
\emph{the exciton composite nature plays a major role}. We assign the
refractive index difference $(n_{+}-n_{-})$ to interactions with the
excited semiconductor, in which the virtual exciton which recombines to restore the
unabsorbed photon, is made of carriers different from the ones of the virtual
exciton coupled to the initial photon.
The linear terms of $(n_{+}-n_{-})$ in the exciton density are the ones we found
in the exciton optical Stark shift, the physics they contain
being identical. We can now go beyond these linear effects and keep a
full control of the physics involved, due to the multiarm ``Shiva''
diagrams for
$N$-body exchanges we have recently introduced. An analytical expression of
$(n_{+}-n_{-})$ is given in terms of the probe photon detuning and the
density of excitons present in the sample (see eqs.(3,4)). Extension of this work to 
more complicated situations such as bulk samples and pump beams with linear
polarization will be done elsewhere.

\begin{figure}[h]
\centerline{\scalebox{0.2}{\includegraphics{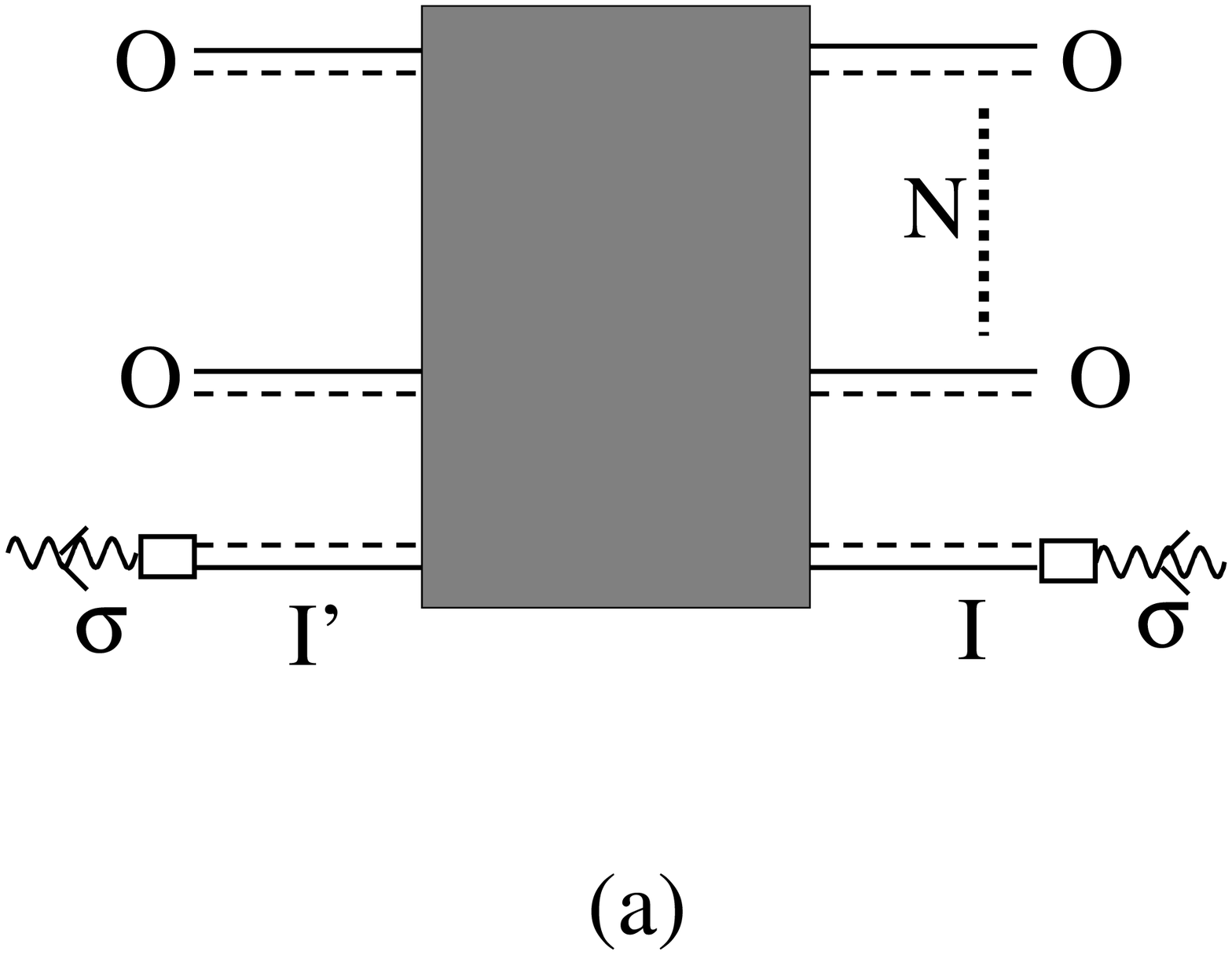}}}
\centerline{\scalebox{0.2}{\includegraphics{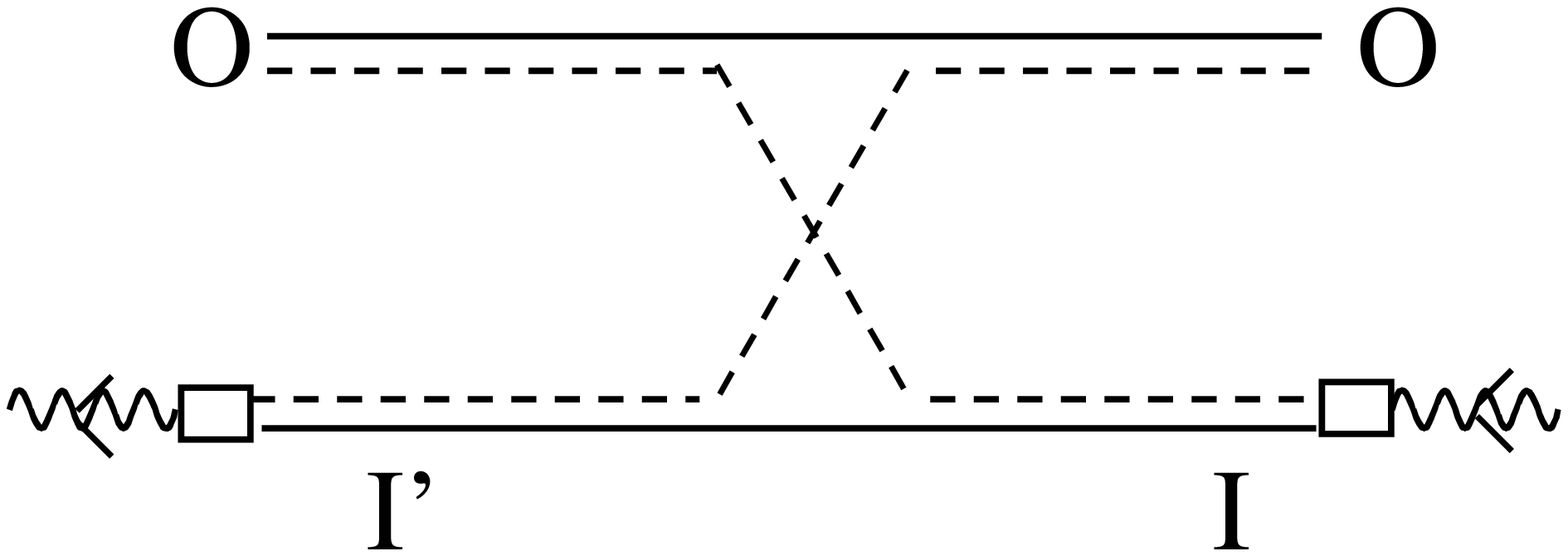}}
\scalebox{0.2}{\includegraphics{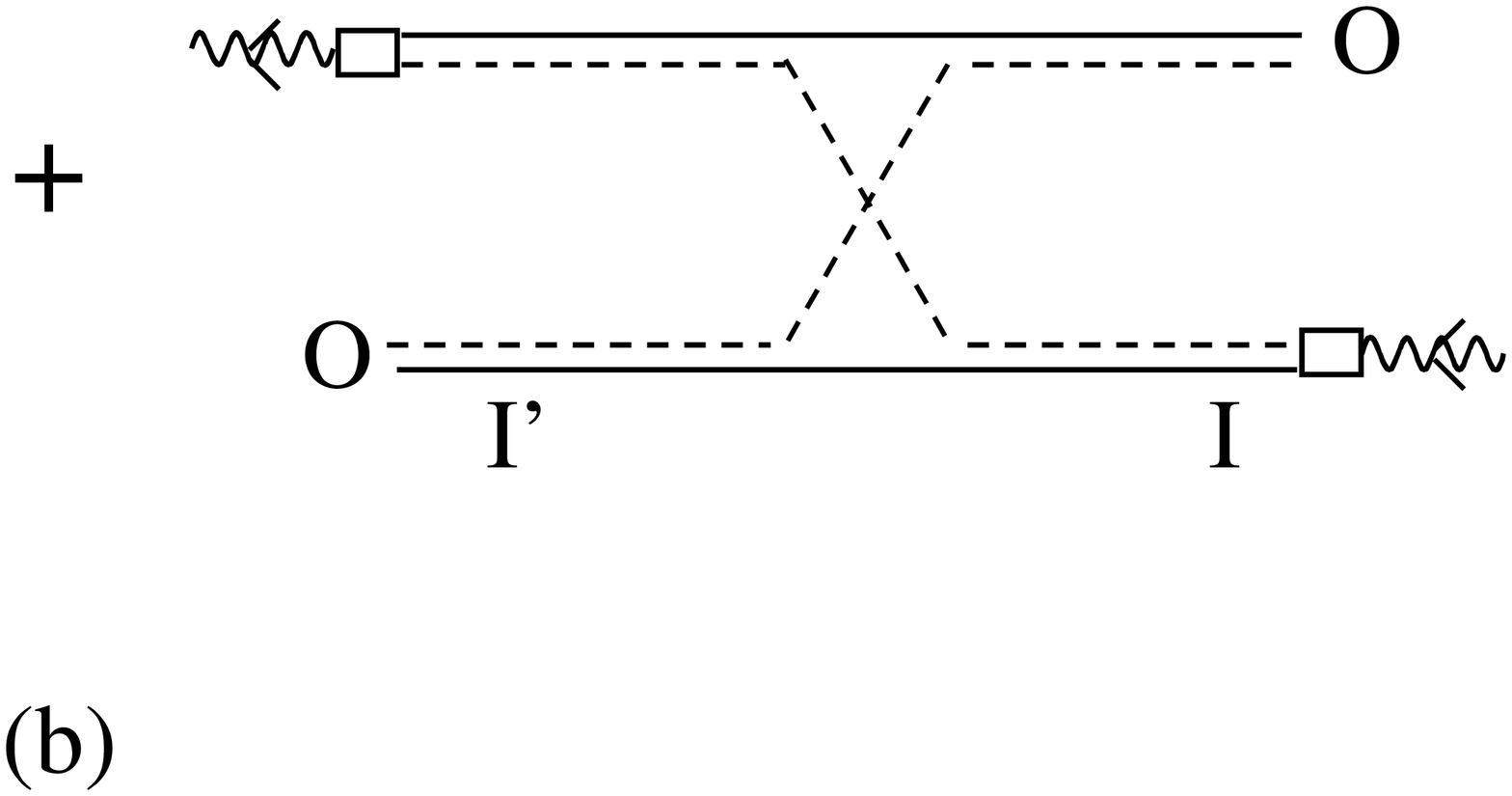}}}
\centerline{\scalebox{0.2}{\includegraphics{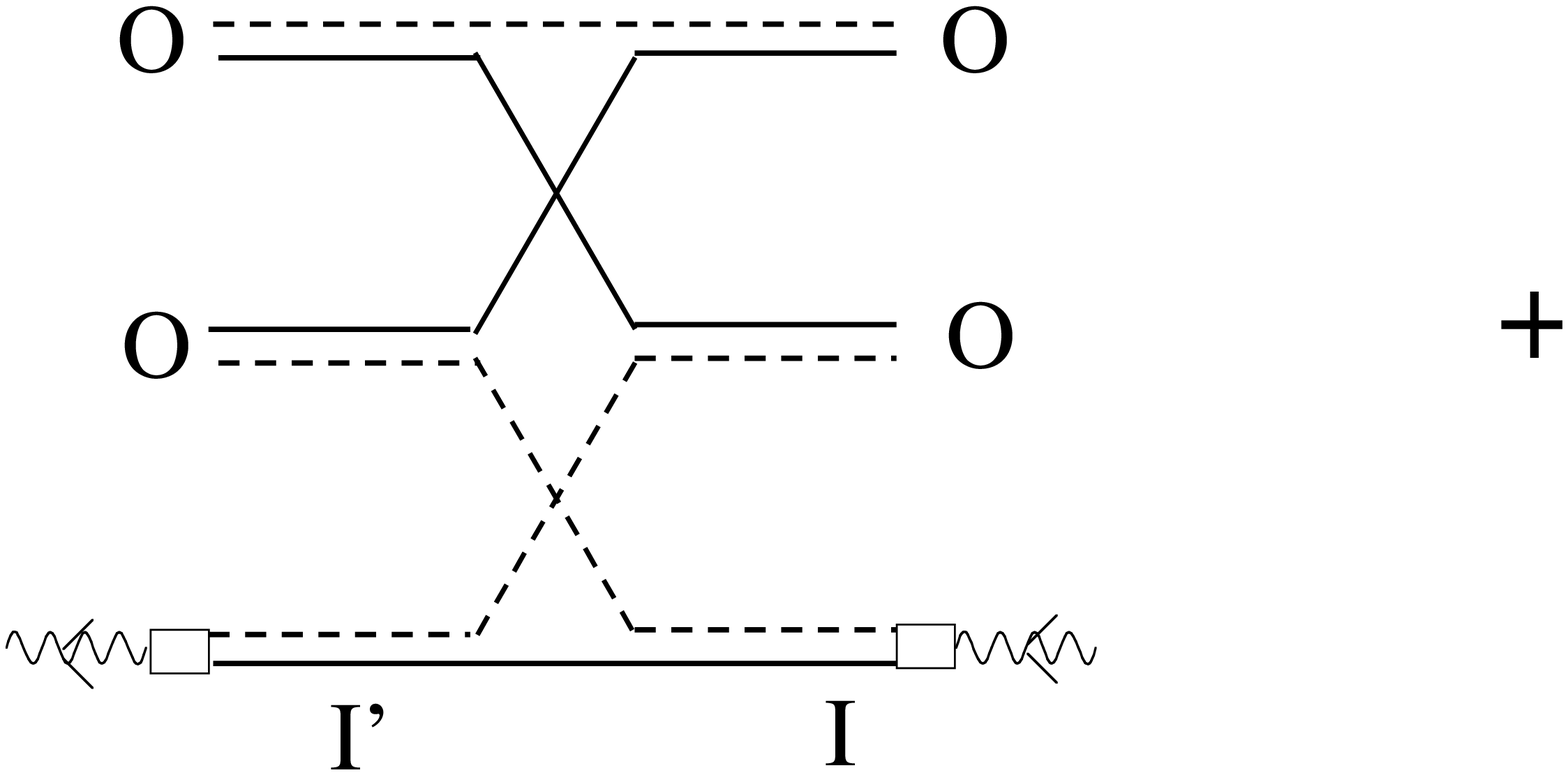}}
\scalebox{0.2}{\includegraphics{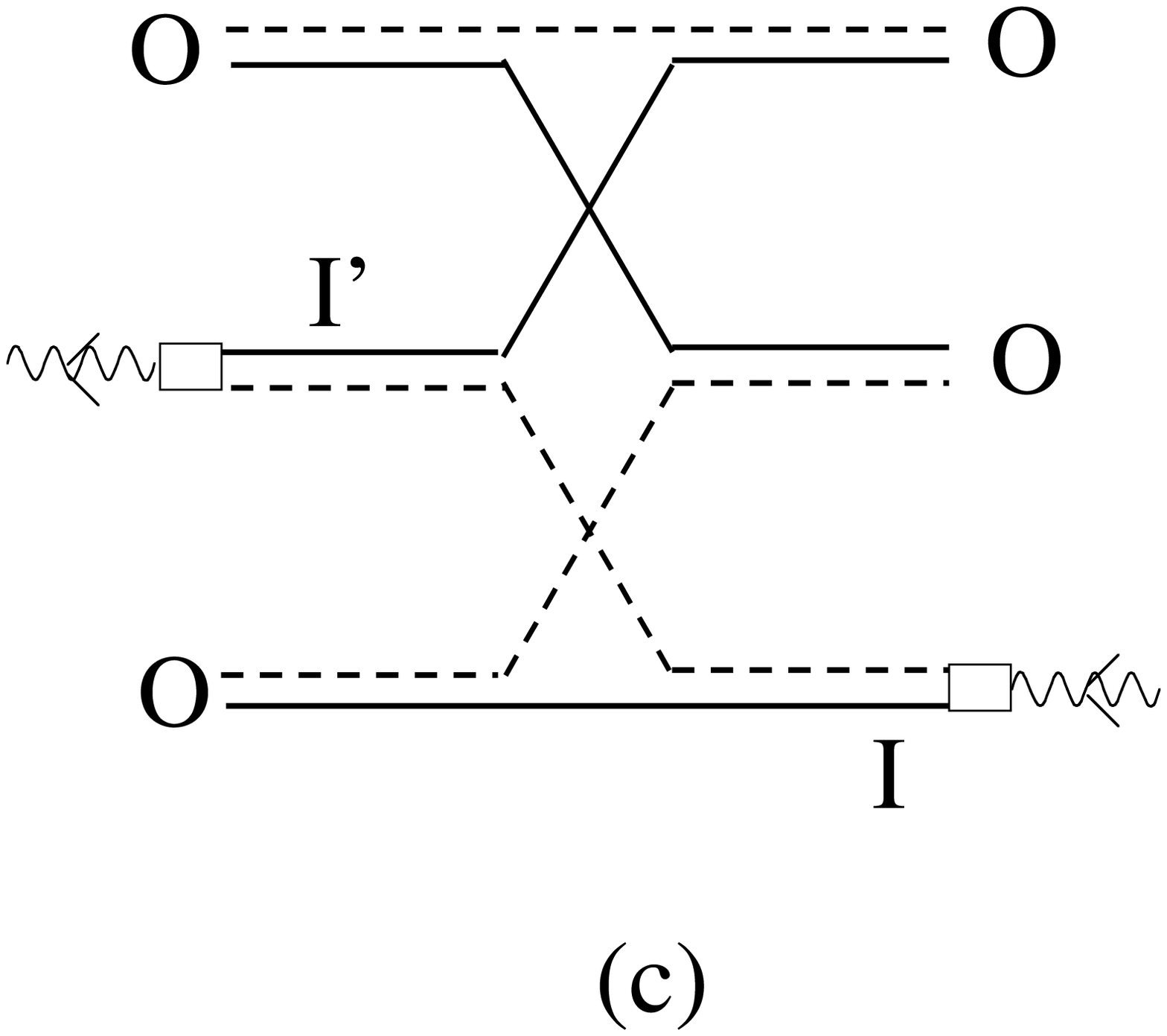}}
\scalebox{0.2}{\includegraphics{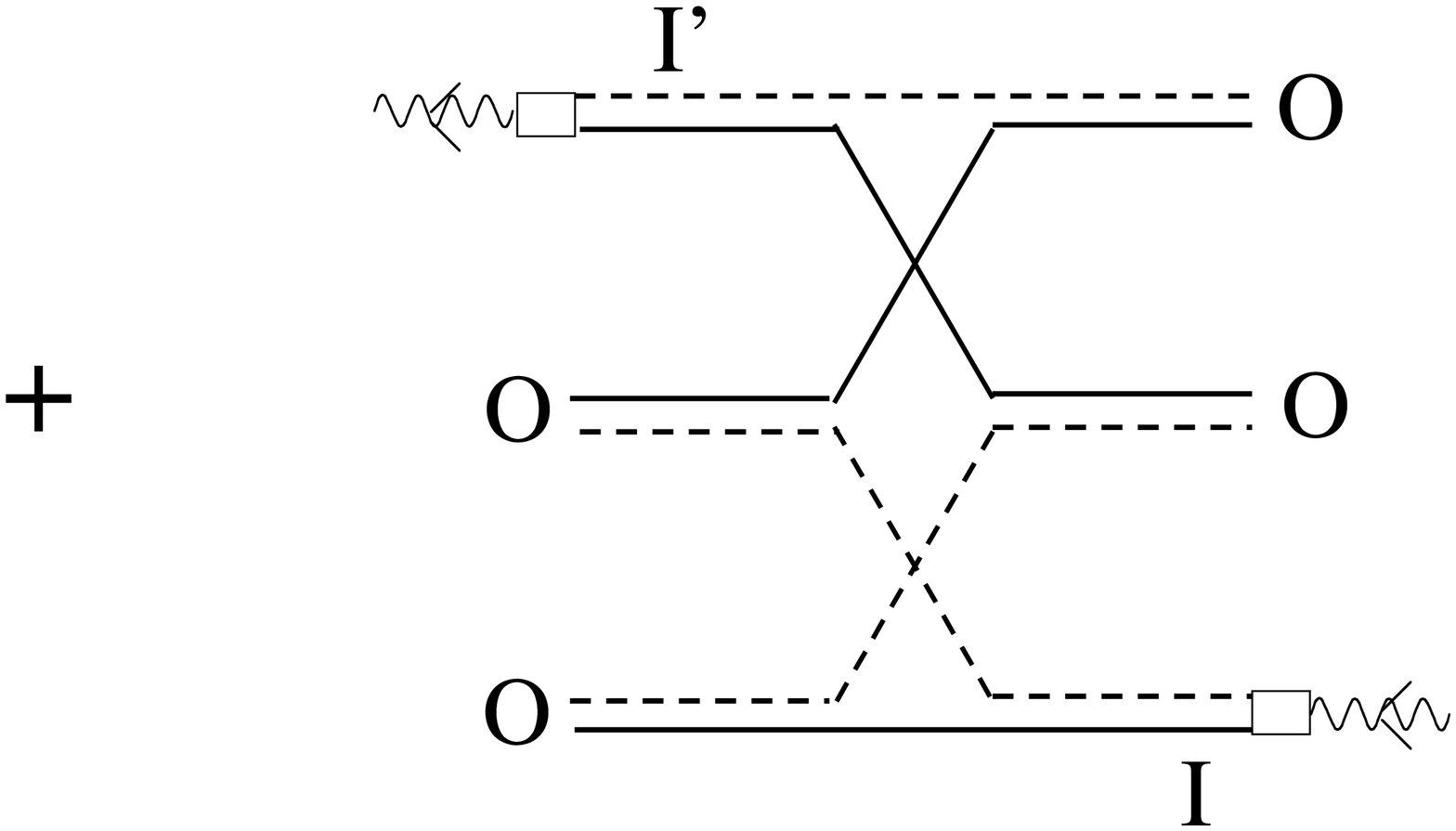}}}
\caption{(a) the ``box'' \emph{a priori} contains any number of Pauli (exchange
without Coulomb) and Coulomb (without exchange) scatterings between the virtual
exciton
$I$ coupled to the
$\sigma$ photon and the $N$ excitons $O$ present in the sample; the only requirement
is to restore the
$N$ excitons $O$. Electrons are represented by solid lines and holes by dashed
lines, the double solid-dashed lines representing the excitons. (b,c) Multiarm
``Shiva'' diagrams for Pauli scatterings between
$I$ and
\emph{one} or
\emph{two} excitons
$O$.
\emph{Faraday rotation is due to processes in which the virtual excitons $I$ and
$I'$ have zero or one common carrier.}}
\end{figure}


\begin{thebibliography}{99}
\bibitem{10}
from the multiarm hindu god. We first called them unproperly ``skeleton diagrams''.
\bibitem{13}
The relative dielectric constant $\epsilon_r=(n+ik)^2$ is related to $S$ through 
$\epsilon_r=\epsilon_{0r}+ S/A^2\omega^2\epsilon_0\mathcal{V}$, where
$\epsilon_{0r}$ is the background contribution, $\epsilon_0$ the vacuum
dielectric constant,
$\omega$ the photon frequency, $\mathcal{V}$ the sample volume. So that
$\mathcal{E}_p=2n_0A^2\omega^2\epsilon_0\mathcal{V}$ for $k$ small and $n_{+}\simeq
n_{-}\simeq n_0=\sqrt{\mathrm{Re}(\epsilon_{0r})}$.
\end{thebibliography}
\end{document}